\documentclass{article}
\usepackage[T1]{fontenc} 
\usepackage[utf8]{inputenc} 
\usepackage{ismir,amsmath,cite,url}
\usepackage{enumitem}
\usepackage{amsmath}
\usepackage{graphicx, tabularx, booktabs}
\usepackage{color}
\usepackage{microtype}
\usepackage{letterspace}
\usepackage{xcolor}
\usepackage{lineno}

\title{From Audio Encoders to Piano Judges: Benchmarking Performance Understanding for Solo Piano}

\multauthor
{Huan Zhang \hspace{1cm} Jinhua Liang \hspace{1cm} Simon Dixon} {
Centre for Digital Music, Queen Mary University of London, UK\\
{\tt\small huan.zhang@qmul.ac.uk}
}

\def\authorname{H. Zhang, J. Liang, S. Dixon}

\begin{document}

\maketitle
\begin{abstract}
Our study investigates an approach for understanding musical performances through the lens of audio encoding models, focusing on the domain of solo Western classical piano music. Compared to composition-level attribute understanding such as key or genre, we identify a knowledge gap in performance-level music understanding, and address three critical tasks: expertise ranking, difficulty estimation, and piano technique detection, introducing a comprehensive Pianism-Labelling Dataset (PLD) for this purpose. 
We leverage pre-trained audio encoders, specifically Jukebox, Audio-MAE, MERT, and DAC, demonstrating varied capabilities in tackling downstream tasks, to explore whether domain-specific fine-tuning enhances capability in capturing performance nuances. 
Our best approach achieved 93.6\% accuracy in expertise ranking, 33.7\% in difficulty estimation, and 46.7\% in technique detection, with Audio-MAE as the overall most effective encoder. Finally, we conducted a case study on Chopin Piano Competition data using trained models for expertise ranking, which highlights the challenge of accurately assessing top-tier performances.

\end{abstract}
\section{Introduction}\label{sec:introduction}

Traditional music understanding tasks focus on composition-level attributes: key, tempo, genre and instrumentation are widely explored \cite{Korzeniowski2017EndNetwork, Pons2018End-to-endScale, FloresGarcia2021LeveragingRecognition}. These attributes are not only tagged individually via end-to-end approaches but have also been the focus of foundation models and various musical representations aimed at learning them in a unified manner\cite{Castellon2021CodifiedRetrieval, McCallum2022SupervisedUnderstanding}, facilitating cross-modal understanding \cite{Gardner2024LLarkMusic, Doh2023LP-MusicCapsCaptioning}.

However, a large portion of human music activity is focused not on the composed songs or pieces themselves, but on the process of learning and performing them \cite{Palmer1997MusicPerformance}. Despite its great importance to the vast community of students, teachers and musicians, the ability to understand performance nuances (challenging techniques, skill varieties, stylistic differences, difficulty grading, etc.) has not been grasped by machines. Sporadic experiments \cite{Ramoneda2022ScoreFingering} of these tasks are conducted, often on small-scale \cite{Parmar2021PianoAssessment, Wang2021Audio-basedMechanism} or proprietary \cite{Seshadri2021ImprovingLearning, Huang2019AutomaticExercises, Morsi2023SoundsPerformances} datasets. Performance understanding, in contrast to the more recognized composition-level music understanding, suffers from scarcity of data \cite{Kim2022OverviewContext, Morsi2024SimulatingLearning}, ambiguity of tasks \cite{Lerch2020AnAnalysis}, and the inherent complexity of modelling and representing expressive elements in performances \cite{Zhang2023SymbolicEvaluation, CancinoChacon2018ComputationalModels}. 

Meanwhile, unified representations and foundation models have advanced several fields by providing robust and versatile frameworks \cite{Gardner2024LLarkMusic, Li2023MutimodalAssistants}, demonstrating their potential to overcome challenges related to data scarcity and task specificity. Building on this precedent and their applications to compositional-level understanding \cite{Won2024FoundationalInformatics}, we extend the capacities of pre-trained audio encoders such as MERT \cite{Li2023MERTTraining} and MULE \cite{McCallum2022SupervisedUnderstanding} into the performance-understanding realm, investigating the shared knowledge between composition- and performance-level understanding: Do pre-trained audio encoders capture performance nuances? Can they categorize performance-related attributes? If not, how can we improve their performance? 

This work is a first step in filling the gaps within the performance understanding realm.  Applying domain-adaptation to pre-trained audio encoders, we work towards a \textit{piano judge} that specializes in ranking performers' skill level,  determining the given repertoire's difficulty and core techniques, thus pursuing a human piano teacher's capability and paving the way to performance understanding in an educational context. Our contributions\footnote{Code available at: https://github.com/anusfoil/PianoJudges}
include:
\begin{enumerate}[nosep]
    \item We benchmark three tasks in the realm of audio performance understanding: expertise ranking, difficulty estimation, and solo piano technique detection.
    \item We leverage four audio representation learners (Jukebox, Audio-MAE, MERT, DAC) and compare their capabilities in tackling the downstream tasks. 
    \item We release Pianism-Labelling Datasets (PLD) with detailed labeling curated for the three tasks, the first large-scale dataset (136 hrs in total) that aims to address performance understanding. 
    \item We fine-tune DAC \cite{Kumar2023High-FidelityRVQGAN} and AudioMAE \cite{Hu2022MaskedListen} by domain-adaptation with solo piano, and compare their performances with pre-trained versions. 
    \item We conduct a case study on Chopin Piano Competition data (\textit{ICPC-2015}), exploring how a trained expertise ranking model can be transferred to rank candidates in the most prestigious competition of the pianistic scene.
\end{enumerate}

\section{Related Work}

\subsection{Performance and education focused understanding}

The exploration of performance through recordings provides rich resources for music understanding. Automatic performance analysis (APA) \cite{Lerch2020AnAnalysis} delves into dimensions from dynamics \cite{Kosta2018DynamicsScore, Kim2023DiffVelModel} to timing \cite{Grachten2009Phase-planeTiming, Shi2021ComputationalMazurkas}, forming the basis for tasks such as performer identification and automatic music performance assessment (MPA)\cite{Kim2022OverviewContext}. The former seeks to attribute performances to their respective musicians based on stylistic and technical signatures \cite{Rafee2021PerformerModels, Zhao2021ViolinistFeatures}, and the latter aims to evaluate the quality and expression of performances. MPA approaches can be further divided by the level of proficiency. For novice players, the emphasis is on technical accuracy, ensuring correct notes and rhythm via score alignment \cite{Sales2023InteractiveEngagement, Morsi2024SimulatingLearning} or detecting conspicuous mistake regions \cite{Morsi2023SoundsPerformances} in a score-free context. Advanced performers are assessed by their expression and musicality, usually in the form of predicting rating scores on multiple dimensions \cite{Zhang2021LearnAssessment, Seshadri2021ImprovingLearning}. Recently, the release of feedback-based assessment data \cite{Matsubara2021CROCUSCritiques, Jiang2023ExpertFeedback} offers the possibility to conduct multimodal MPA in a more personalized manner.

On the other hand, the analysis of performances in an educational context emphasizes the identification of challenges and learning opportunities within the repertoire: expert-annotated difficulty level is predicted from symbolic scores \cite{Ramoneda2023CombiningClassification} via a machine learning classification approach that merges musicologically-inspired score features. At a more granular level, we would like to identify instrument-specific techniques that demand practice. For example, techniques such as \textit{acciacatura} and \textit{portamento} on Chinese bamboo flute can be identified from spectro-temporal patterns \cite{Wang2020PlayingScattering}, but similar problems have yet to be explored on piano because of the homogeneity of piano sound. Other learning aid information such as fingering \cite{Ramoneda2022ScoreFingering, Srivatsan2022ChecklistPrediction} and bowing \cite{Hall2011CalibratingStudents} can also be predicted. In this work, we focus on expertise, difficulty and technique estimation by extracting relevant information from performance audio. 
We are the first of its kind in piano technique detection and expertise ranking, and an audio representation approach in compare with the current difficulty estimation \cite{Ramoneda2024CanDataset, Ramoneda2022ScoreFingering}.



\subsection{Leveraging audio representations for downstream tasks}
\label{subsec:lev_audiorep}

The surge of learning audio representations was originally motivated by the generative models such as AudioLM \cite{Borsos2023AudioLMGeneration} and MusicLM \cite{Agostinelli2023MusicLMText}. 
Jukebox \cite{Dhariwal2020JukeboxMusic} is a generative model trained on 1.2M songs. Subsequent work \cite{Castellon2021CodifiedRetrieval, Manilow2022SourceModels} has shown that Jukebox’s representations can be effective features for task-specific linear classifiers. Jukebox embeddings have also been employed in multimodal learning \cite{Gardner2024LLarkMusic} of music captioning and reasoning tasks. 
MERT \cite{Li2023MERTTraining} uses masked language modelling (MLM) style acoustic self-supervised pre-training. With a music teacher and an acoustic teacher, MERT demonstrates good performance in downstream music understanding tasks and extends its music understanding ability into question answering and captioning \cite{Liu2024MusicCaptioning} by generating music representations to aid language model. 

Audio-MAE \cite{Hu2022MaskedListen} is a vanilla 12-layer transformer that learns to reconstruct randomly-masked spectrogram patches. The output feature map from the penultimate block of an Audio-MAE encoder has been used to encode fine-grained patterns in audio \cite{Liang2023AcousticCapabilities}.  
Different from previous approaches, Descript-Audio-Codec (DAC) \cite{Kumar2023High-FidelityRVQGAN} is a neural audio compression autoencoder that compresses high-dimensional signals into lower dimensional discrete tokens. DAC has been proven useful in a generative context \cite{Garcia2023VampNetModeling}, but there have been few attempts to explore it with downstream understanding tasks \cite{Puvvada2024DiscreteRecognition}. 

The four aforementioned audio representations are chosen for our investigation. Since they are constructed from different theoretical approaches (quantized codecs vs.\ continuous spectrograms) and trained on different data (general audio vs.\ music), this variety presents an opportunity to evaluate the extent to which the encoded information contributes to performance understanding. 

\section{Methodology}

\subsection{Downstream problem definitions}

\subsubsection{Expertise ranking}
\label{subsec:expertise_ranking}

We formulate our assessment into a ranking problem: given audio performances $p_1$ and $p_2$, which one has the higher expertise?  We define three coarse levels of expertise (beginner, advanced and virtuoso), represented by integers 0, 1 and 2, respectively, and define a function $Q$ which maps a performance to one of these levels. Instead of directly predicting the absolute expertise level $Q$, we learn a 2-way or 4-way ranking function between each pair of performances from different levels, $R_2$ or $R_4$, as below:

{\tiny
\begin{align}
R_2 &= 
\begin{cases} 
      0 & Q(p_1) < Q(p_2) \\
      1 & Q(p_1) > Q(p_2)
\end{cases}
&
R_4 &= 
\begin{cases} 
      0 & Q(p_2) - Q(p_1) = 2 \\
      1 & Q(p_2) - Q(p_1) = 1 \\
      2 & Q(p_1) - Q(p_2) = 1 \\
      3 & Q(p_1) - Q(p_2) = 2
\end{cases}
\label{eq:rank_definition}
\end{align}
}%

The motivation behind is to teach the model a relative notion of expertise, instead of an absolute level or category of the performance quality. In real life and competition settings (as will be discussed in Sec~\ref{sec:competiton_winner}), we are more interested in the comparative skill level among a set of candidates. 

\subsubsection{Difficulty estimation}
\label{subsubsec:difficulty}

Following the literature \cite{Ramoneda2023CombiningClassification, Ramoneda2022ScoreFingering, Zhang2023SymbolicEvaluation} on difficulty level prediction, we formulate the problem as a classification task with 9 difficulty classes, given the dataset described in Section~\ref{subsec:difficulty}, which has 9 levels of difficulty annotation. Given that the difficulty annotation is subjective and boundaries between levels are fuzzy, we also report the results of 3-class estimation by merging the level groups, as in \cite{Ramoneda2023CombiningClassification}.

\subsubsection{Technique identification}
\label{subsec:technique_identification}

Given a piece, a piano teacher can immediately identify the most challenging passage(s) that would require students hours of practice to master: intense octave runs, fast flowing scales, repeating notes that require finger iteration, etc. 

In the technique-specific dataset (Section~\ref{subsec:techniques_data}), we include 7 common techniques 
and formulate a multi-label classification task for technique identification. Given that our labels are relatively sparse, we also experiment with the case of single-label prediction in which predicting any one of the multiple labels is considered correct. 

\begin{table}[t]
    \centering
    \begin{tabular}{c|c|c|c}
        \textbf{Encoder} & \textbf{C} & \textbf{F (Hz)} & \textbf{Dim} \\ \hline
        Jukebox & 2048 & 345 & 64 \\
        MERT    & - &  75  & 1024 \\
        Audio-MAE & - & 51.2 & 768 \\
        DAC & 9$\times$1024 & 87 & 1024 \\
        Spectrogram & - & 150 & 128 \\
    \end{tabular}
    \caption{Specifications of the audio encoders as well as the spectrogram baseline: C is the codebook size, F is the frame rate in Hz and Dim is the hidden dimension of the embedding (mel-bins for spectrogram).}
    \label{tab:audio_encoders}
\end{table}

\subsection{Audio embeddings and encoder fine-tuning}

An overview of the used audio embeddings is given in Table~\ref{tab:audio_encoders}. For Jukebox, we employ the 345 Hz sample rate encoding, and for Audio-MAE, the 768-dimensional embedding is taken from the ViT-B Transformer encoder. Additionally, we considered a spectrogram baseline, as a low-level representation to compare with the trained embeddings. We use 128 mel bins, an FFT of 400 samples, and hop size of 160 samples, resulting in a spectrogram with frame rate of 150 Hz and 128 dimensions that feeds into the prediction head module like the trained embeddings. 

We examine whether fine-tuning the two generic-audio-trained encoders DAC and Audio-MAE with domain-specific data results in a performance boost
. The two encoders are fine-tuned using their original self-supervision objective on around 2k hours of solo piano recordings, from datasets of MAESTRO \cite{Hawthorne2018EnablingDataset}, ATEPP \cite{Zhang2022ATEPPPerformance}, SMD \cite{Konz2011SaarlandData}, Mazurkas\footnote{\url{http://www.charm.rhul.ac.uk/index.html}} as well as the novel PLD data introduced in this work. For DAC, the fine-tuning lasts for 25k iterations while the Audio-MAE is fine-tuned for 64 epochs. 

\subsection{Experiments}

\begin{figure}[t]
    \centering
    \includegraphics[trim={2cm 3cm 2cm 2.2cm},clip,width=\linewidth]{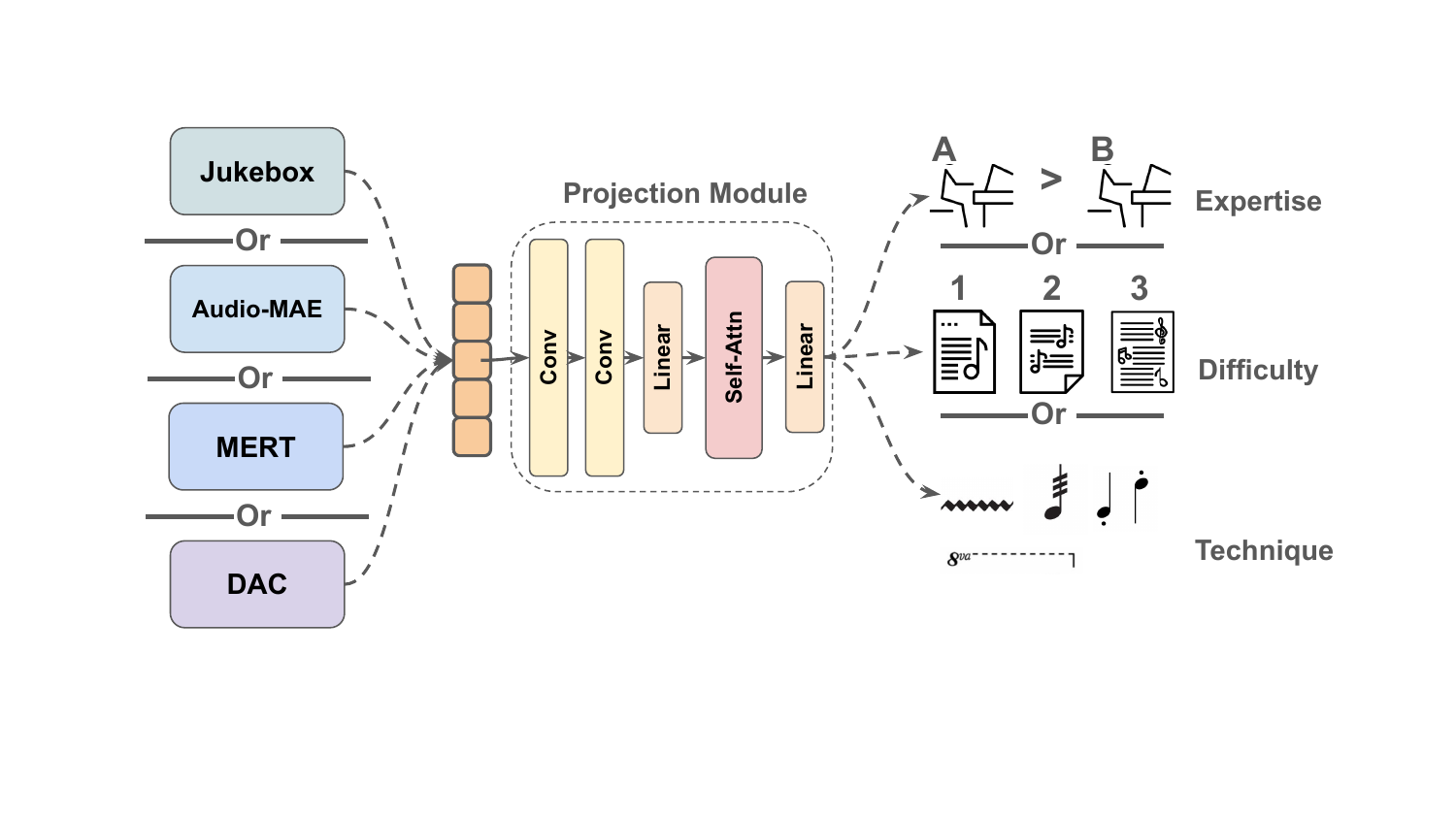}
    \caption{Overview of our tasks and experiment pipeline.For the expertise ranking, two audio embeddings are concatenated in the time dimension.}
    \label{fig:pj_pipeline}
\end{figure}

For all encoders, we first compute 10-second segment audio embeddings (or spectrograms), and include a maximum of 5 minutes (30 segments) of audio as input with padding. The concatenated embedding of each audio track is of shape $(30, F\times10, D)$ where $F$ is the frame rate and $D$ is the embedding dimension as shown in Table~\ref{tab:audio_encoders}.

Given the audio embedding, 
we transform it through a prediction head module that consists of two 2D convolutional layers ($n_\mathit{kernel}=7, n_\mathit{stride}=5$), one linear layer to align different input dimensions, and one self-attention layer ($n_\mathit{heads}=2$, $d=128$), followed by a final linear layer that projects to the desired classes of each task. A full pipeline of the experiments is shown in Figure~\ref{fig:pj_pipeline}. As is standard practice\cite{Manilow2022SourceModels, Castellon2021CodifiedRetrieval}, we maintain a straightforward projection module design, aiming to minimize its influence on the probing performance. 

Regarding each individual task, we run a grid search on the hyper-parameters for learning rate, weight decay, batch size, etc. The details for the final training parameters for each task are documented in the project page\footnote{\url{https://bit.ly/3SYzozY}}. All fine-tuning and training are conducted on one NVIDIA A5000.


\section{Pianism-labeling dataset}

The pianism-labeling dataset (PLD) includes audio and annotations for three notions that are centrally relevant to pianism and piano education: \textbf{expertise}, piece \textbf{technique} and \textbf{difficulty}, where dataset statistics are specified in Table~\ref{tab:dataset_statistics}.  All of the labeling, metadata correspondence, as well as examples are available on the project page. 

\begin{table}[t]
    \centering
    \begin{tabular}{c|cccc}
         & task type &  classes &  tracks & len. (s) \\ \hline
        \textbf{Expertise} & Multi-class & 2 or 4 & 1694 & 167.4 \\
        \textit{ICPC-2015} & Multi-class & 2  & 137  & 1827.0 \\ \hline
        \textbf{Difficulty} & Multi-class & 3 or 9 & 737 & 269.8 \\ \hline
        \textbf{Techniques} & Multi-label & 7 & 222 & 45.5 \\
    \end{tabular}
    \caption{Dataset statistics (number of classes, number of tracks and average duration) for each task.}
    \label{tab:dataset_statistics}
\end{table}

\subsection{Expertise}
\label{subsec:expertise}

We curated a collection of solo piano recordings from YouTube, each annotated with an expertise level. Their categorization was based on information gleaned from the YouTube channels' descriptions, which provided insights into the background of the recordings. This categorization process was validated by two college-level piano students to ensure accuracy. 
\begin{itemize}[nosep]
    \item \textbf{Beginner} (562): Amateur level, featuring mostly adult self-taught learners' practice recordings. 
    \item \textbf{Advanced} (570): Performances of music students and junior competition recordings. 
    \item \textbf{Virtuoso} (562): Famous pianists' recordings sourced from the ATEPP \cite{Zhang2022ATEPPPerformance} dataset. To balance with the other groups, we randomly select a subset of 562 of the 11K recordings.
\end{itemize}

The repertoire of selected performances is mainly focused on the Western Classical repertoire, with some rearranged folk and pop songs at the Beginner level. Indeed, it is challenging to align the performed repertoire across levels since the complexity of played pieces increases with the expertise: e.g. Beginners' pieces are shorter (av.\ 128.9 s) than Advanced (201.1 s) or Virtuoso (171.8 s) tracks. 

In experiments, the three levels are first individually split into train and test subsets and then paired up randomly. Each recording only shows up once in the pairs to prevent leakage, which results in 2694 pairs in training.

\subsubsection{ICPC-2015}
\label{subsubsec:icpc_2015}

In this task we aim to assess whether the learnt comparative ranking objective can be applied to the professional domain, the International Chopin Piano Competition, with data gathered from the 2015 edition (\textit{ICPC-2015 dataset})\footnote{\url{https://github.com/cyrta/ICPC2015-dataset}}. We employ only the preliminary round performances to ensure limits on the length and instrumentation (i.e.\ solo), and assume that the overall better players clearly demonstrate their skills in the preliminary round performance. Out of 160 candidates, 137 recordings are successfully retrieved. 

We compile the data into ranking pairs similar to Section~\ref{subsec:expertise}, by first assigning a score $S(c)$ for each candidate based on their progression into the following rounds. For every round that the candidate passes into, the score is incremented with 1 point. For candidates $a, b$ with their respective scores $S(a), S(b)$, all preliminary round recordings are formed into pairs with ranking as in Eq.~\ref{eq:rank_definition}. 
As shown in Table~\ref{tab:dataset_statistics}, the preliminary round recordings have an average duration of 30 minutes. Thus, we obtain paired ranking results for each pair of 5-minute segments (30 segments in total) and use majority voting to obtain the final rank among two recordings.

\subsection{Difficulty}
\label{subsec:difficulty}

We employ the \textit{Can I Play It?} (CIPI)\cite{Ramoneda2023CombiningClassification} dataset for our task of difficulty prediction. Given that the original dataset is sourced in symbolic MusicXML, we obtain the performance audio from YouTube by querying the metadata followed by manual correction to enforce piece alignment. Note that the performances are sourced from different levels of playing rather than virtuoso recordings only, with the aim of learning a more general view of audio difficulty. In the CIPI dataset, difficulty labels are annotated by Henle Verlag\footnote{\url{https://www.henle.de/}}, a renowned publisher in the music education community. The ratings range from 1-9 and span 29 composers. Note that we split the movements from sonata or other multi-movement compositions, resulting in the 737 audio tracks shown in Table~\ref{tab:dataset_statistics} compared to 637 compositions in the original metadata. We also use the same train-test split as the original dataset.

\subsection{Techniques}
\label{subsec:techniques_data}

The technique dataset contains 222 recordings with an average duration of 45 seconds, demonstrating one or more canonical piano techniques from seven categories taken from piano practice literature \cite{Sandor1981OnExpression}. The excerpts are taken from etude books like Beyer or Czerney, or passages from performance repertoire (e.g.\ dense octave run from \textit{Chopin op.25 no.10}).  Besides YouTube sourcing, 41 out of 222 recordings are recorded by the authors, if the specific passages containing the techniques are not publicly available in any recording. The categories of techniques are:
\begin{itemize}[nosep]
    \item \textbf{Scales} (48): Pure scale run across octaves. Can be both hands or one hand.
    \item \textbf{Arpeggios} (40): Pure arpeggio run across octaves, or music passages that are accompanied with arpeggios, or melody that is constructed on arpeggiated chords.
    \item \textbf{Ornaments} (31): Including grace notes, trills, \textit{mordents}, \textit{acciacatura}. Note that we do not balance these subclasses, and the most common ornament in our samples is grace note. 
    \item \textbf{Repeated notes} (35): Musical passages that feature a series of repeated single notes. 
    \item \textbf{Double notes} (36): Musical passages that feature sequences of simultaneous intervals (mostly thirds, but also fourths and sixths), where the intervals are performed with one hand. 
    \item \textbf{Octave} (35): We differentiate octaves from double notes because of their sheer importance in piano repertoire, as well as its distinctive sonority. 
    \item \textbf{Staccato} (41): Musical passages that are predominately performed by \textit{staccato} articulation. 
\end{itemize}

We formulate the prediction task as multi-label classification since a musical passage is often associated with multiple techniques. Among the 222 recordings, we have 40 labeled with two techniques and two recordings with three techniques. Note that besides scales and arpeggios, few other techniques exist in their pure form (i.e.\ an entire music passage of trills). Thus we aim to identify the most prominent technique present in the recording.

\section{Results}

\subsection{Expertise Ranking}
\label{subsec:expertise_results}

We train the projection module in 2-way and 4-way ranking as described in Section~\ref{subsec:expertise_ranking}, and show results in Table~\ref{tab:main_result} (left). For 2-way ranking, we achieve up to 93.56\% accuracy, indicating a clear distinction between recordings of varying levels of expertise in most cases. Audio-MAE outperforms the other three audio encoders while Jukebox embeddings contain the least information for discerning the level of playing. The result of 4-way prediction is similar with Audio-MAE performing the best with 84\% accuracy, indicating a good capability to distinguish larger expertise differences (beginner vs.\ virtuoso) from smaller ones.  The baseline spectrogram achieves much lower metrics on both classifications, indicating that the pre-trained encoders capture more relevant nuances of musical performance. However, we are also aware that the three levels of data differs not only on performance but also on repertoire and recording environment. The effect of fine-tuning with solo piano domain data is not salient in this task: the fine-tuned Audio-MAE achieved roughly the same performance while DAC actually declined. 

\begin{table*}[ht]
\small
\centering
\label{tab:merged_results}
\begin{tabular}{@{}lcccccccccccccc@{}}
\hline
\hline
 & \multicolumn{4}{c}{\textit{Expertise Ranking}} & \multicolumn{4}{c}{\textit{Difficulty Estimation}} & \multicolumn{6}{c}{\textit{Technique Identification}} \\
 \cmidrule(r){2-5} \cmidrule(r){6-9} \cmidrule(r){10-14} 
 & \multicolumn{2}{c}{2-way} & \multicolumn{2}{c}{4-way} & \multicolumn{2}{c}{9-way} & \multicolumn{2}{c}{3-way} & \multicolumn{3}{c}{Multi} & \multicolumn{3}{c}{Single} \\
 & \textit{Acc} & \textit{F1} & \textit{Acc} & \textit{F1} & $Acc_0$ & $Acc_1$ & $Acc_0$ & $F1$ & $mAP$ & $AUC$ & $Acc$ & $Acc$ & $F1$ & \\[-.5ex]
 \cmidrule(r){1-5} \cmidrule(r){6-9} \cmidrule(r){10-14} 
\textit{pre-trained} & & & & & & & & & & & & & & \\[-1ex]
\cmidrule(r){1-5} \cmidrule(r){6-9} \cmidrule(r){10-14} 
Spec & 75.90 & 74.73 & 52.34 & 49.94 & 32.98 & 59.17 & 67.21 & 66.75 & 57.49 & \textbf{71.13} & 73.02 & \textbf{46.67} & \textbf{39.26} & \\
Jukebox & 84.51 & 83.79 & 60.41 & 56.75 & 33.41 & 55.36 & 60.49 & 58.27 & 49.33 & 59.79 & 73.33 & 25.44 & 23.53 & \\
Audio-MAE & 93.48 & \textbf{92.84} & \textbf{84.21} & \textbf{81.20} & 31.60 & \textbf{66.09} & \textbf{79.03} & 75.11 & 60.69 & 67.51 & 77.46 & 42.22 & 39.81 & \\
MERT & 89.48 & 88.73 & 82.12 & 78.81 & 26.55 & 62.28 & 73.13 & 71.38 & 55.76 & 69.05 & \textbf{79.37} & 37.78 & 35.85 & \\
DAC & 86.84 & 87.91 & 77.77 & 76.24 & 27.61 & 59.86 & 69.64 & 69.87 & 48.50 & 57.87 & 78.73 & 24.44 & 23.61 & \\[-.5ex]
\cmidrule(r){1-5} \cmidrule(r){6-9} \cmidrule(r){10-14} 
\textit{fine-tuned} & & & & & & & & & & & & & & \\[-1ex]
\cmidrule(r){1-5} \cmidrule(r){6-9} \cmidrule(r){10-14} 
Audio-MAE & \textbf{93.56} & 90.22 & 82.26 & 77.82 & \textbf{33.67} & 60.21 & 77.73 & \textbf{75.84} & \textbf{61.81} & 67.61 & 79.05 & 35.56 & 33.73 & \\
DAC & 82.87 & 81.83 & 78.41 & 76.23 & 28.63 & 61.59 & 64.45 & 62.34 & 50.77 & 59.80 & 79.68 & 26.67 & 25.66 & \\
\hline
\hline
\end{tabular}
\caption{From left to right: results of 2-way and 4-way expertise ranking, 9-way and 3-way difficulty estimation, multi-label and single-label technique prediction. Best results are highlighted in bold.}
\label{tab:main_result}
\end{table*}


\subsubsection{Discussion: How far are we from predicting the Chopin Competition Winner?}
\label{sec:competiton_winner}

From the trained 2-way expertise ranking module, we apply the \textit{ICPC-2015} pairs as a testing set as described in Section~\ref{subsubsec:icpc_2015}. Ideally, the model should discern the three levels of piano expertise by identifying specific nuances in performance that distinguish, for example, virtuosi from advanced students. Such insights could then be applicable to evaluating competition-level performances.
In Table~\ref{tab:competition_result}, ``fitting'' indicates that we first fit the trained model on half of the candidates' pairs for 5 epochs and test on the other half. Without fitting, we only evaluate on these same testing pairs using the model trained in Section~\ref{subsec:expertise_results}.

\begin{table}[h!]
\small
\centering
\begin{tabular}{@{}lcccc@{}}
\toprule
 & \multicolumn{2}{c}{w/o. fitting} & \multicolumn{2}{c}{w. fitting} \\
\cmidrule(r){2-3} \cmidrule(l){4-5}
 & \textit{Acc} & \textit{F1} & \textit{Acc} & \textit{F1} \\[-.5ex]
\midrule
\textit{pre-trained} & & & &  \\[-1ex]
\midrule
Spec & 52.91 & 52.79 & 49.27 & 49.04 \\
Jukebox & 46.63 & 44.92 & 48.27 & 47.13 \\
Audio-MAE & 56.86 & 55.86 & 59.08 & 58.76 \\
MERT & 49.07 & 46.78 & 53.05 & 52.54 \\
DAC & 42.17 & 41.71 & 53.67 & 53.67 \\[-.5ex]
\midrule
\textit{fine-tuned} & & & &  \\[-1ex]
\midrule
Audio-MAE & 54.32 & 50.14 & 54.89 & 49.81 \\
DAC & \textbf{60.49} & \textbf{60.27} & \textbf{59.87} & \textbf{59.84} \\
\bottomrule
\end{tabular}
\caption{2-way paired-ranking test result for the competition dataset \textit{ICPC-2015}. }
\label{tab:competition_result}
\end{table}

Several interesting observations are made from this experiment: 1) Transferring the learnt expertise ranking into assessing competition-level playing (which shall all belongs to the virtuoso tier within our training) is challenging, considering random guess baseline of 50\% accuracy in predicting the better performer within a pair. The best we achieve is slightly above 60\%, possibly because that the outcomes of competitions often transcend mere audio content to include performative expression like motions, resulting in a \textit{sight over sound} phenomenon \cite{Tsay2013SightPerformance}.  2) Adaptation on the competition set does not significantly boost the performance. For the pre-trained embeddings the accuracies slightly increase after fitting, but it has no effects on the fine-tuned embeddings. 3) The fine-tuned DAC embeddings, despite having a lower performance in the ranking task with three levels, largely outperform other models in ranking the candidates in a competition setting. 

\begin{figure}[!h]
    \centering
    \includegraphics[width=\linewidth]{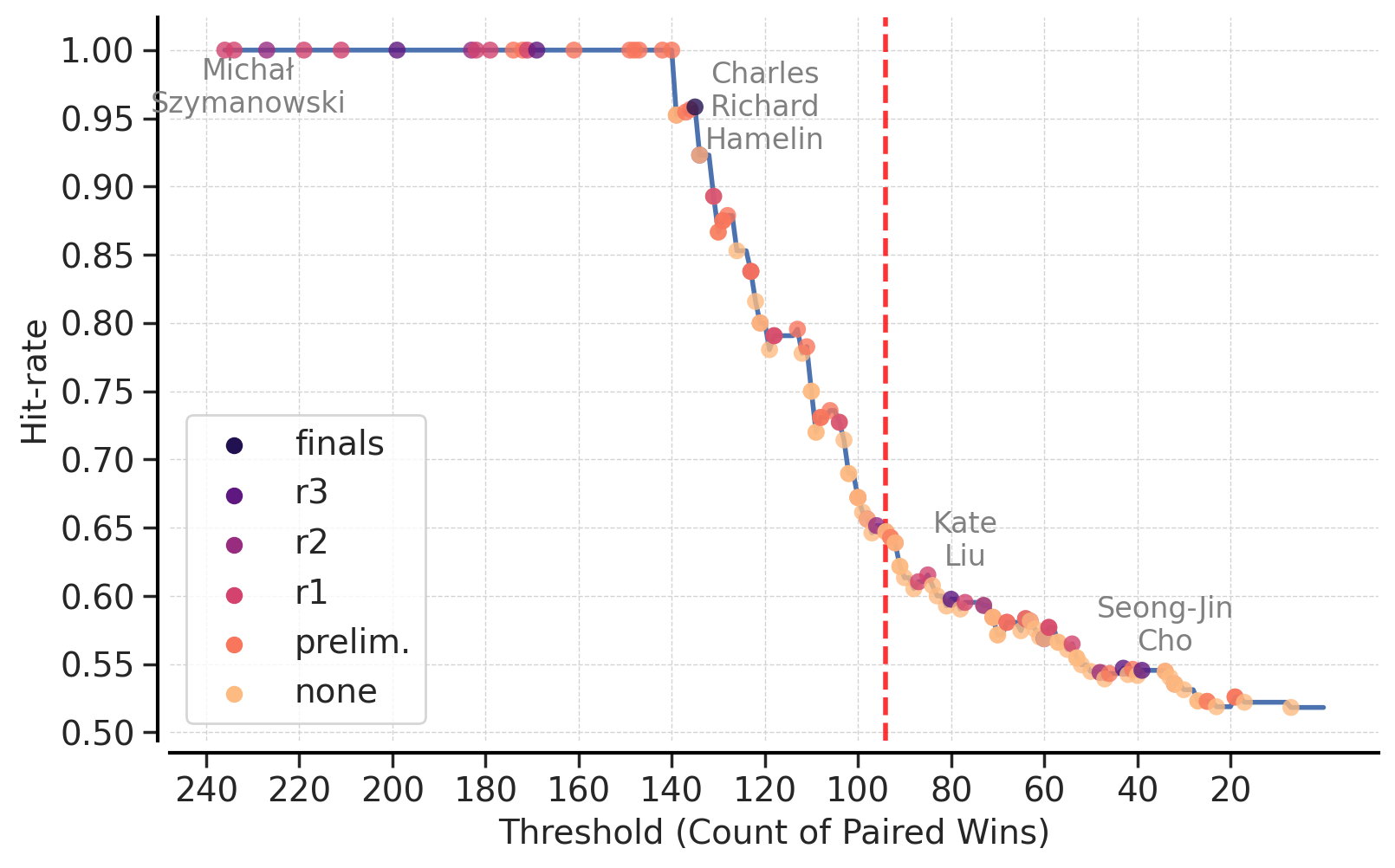}
    \caption{Paired win count threshold vs. hit-rate for preliminary round pass prediction. Each candidate is a data point colored by their ground-truth result tier (i.e.\ the highest round they progressed from). The red dashed line is the cut-off of half the candidates.}
    \vspace{-1.5em}
    \label{fig:pass_predict}
\end{figure}

Using the paired prediction results from best model (fine-tuned DAC w/o. fitting), we translate pair-wise predictions into a global ranking. Each candidate is ranked by how many wins they obtain in the `paired matches'. Each candidate is involved in 272 pairs, given 137 candidates and we infer on each pair (136) and its inverse. Figure~\ref{fig:pass_predict} shows the relationship between our predicted candidate win counts and the preliminary round pass hit-rate (i.e.\ what proportion of candidates actually passed the preliminary round). \textit{Michał Szymanowski} is the predicted best candidate who wins in the most pairs. Overall, there exists a good correlation between our predicted win counts rank and candidates' ground truth performance: the top 18 predicted candidates all passed the preliminary round, with many of them progressing into round 2 or 3 (demonstrated by the color in Figure~\ref{fig:pass_predict}). Down to the cut-off threshold of half of the candidates, 65\% of them passed the preliminary round. Finalists, however, are not necessarily predicted accurately: the winner \textit{Seong-Jin Cho} only ``wins'' 39 matches and is placed towards the end in this rank, as is the third placed \textit{Kate Liu}. Only \textit{Charles Hamelin} (2nd place) and \textit{Dmitry Shishkin} (6th) are placed relatively high in our ranking.


\subsection{Difficulty estimation}


The difficulty estimation experiments are defined in Section~\ref{subsubsec:difficulty} on the \textit{CIPI} dataset. In Table~\ref{tab:main_result} (middle), we report the \textit{accuracy within n} ($\text{Acc}_n$) for the 9 class prediction. Defined as Eq.~\ref{eq:accuracy}, $\text{Acc}_n$ aligns with the ordinal nature of the task and we observe results for $n = 0$ (exact match) and $n = 1$ (allowing for one-class deviation). 
\vspace{-1mm}
\begin{equation}
\small
\label{eq:accuracy}
\text{Acc}_n = \frac{1}{|C|} \sum_{c \in C} \frac{|\{ y \in S_c : |\hat{f}(x) - c| \leq n\}|}{|S_c|}
\end{equation}
\vspace*{-2mm}

The Audio-MAE embeddings yield overall the best performance for both 9-way and 3-way estimation. But it is worth noting that the untrained spectrogram baseline actually achieved accuracy metrics on-par with the audio encoders (32.98\% vs.\  33.67\% in 9-way estimation), even higher than the worst-performing embedding of DAC.

The best we achieve with 9-class $\text{Acc}_0$ is 33.67\% (compared to the same-set symbolic data baseline \cite{Ramoneda2023CombiningClassification} of 39.47\%). However, this is based on the fact that our audio embeddings are capped to 5 minutes, removing the effect of the major feature of piece length. For the 3-way classification we achieved accuracy that is on-par with the symbolic baseline, with the best $\text{Acc}$ of 79.03\%, demonstrating that the complexity of piano repertoire can also be encoded with the current pre-trained representation. 

Interestingly, the $\text{Acc}_0$ and $\text{Acc}_1$ metrics do not improve hand-in-hand: Jukebox embeddings achieved the highest $\text{Acc}_0$ among the pre-trained models, but performed worst on $\text{Acc}_1$ since its prediction is sparse and scattered from observing the confusion matrix. The fine-tuned models exhibit a modest enhancement in performance metrics, as in the $\text{Acc}_0$ for Audio-MAE and 3-way \textit{F1} for DAC, also helped with better generalization and less overfitting.

\subsection{Technique identification}


The technique identification experiment is performed as both multi-label and single-label prediction, as formulated in Section~\ref{subsec:technique_identification}.
In Table~\ref{tab:main_result} (right), we report the mean-Averaged-Precision ($mAP$) and Area Under the Receiver Operating Characteristic Curve ($AUC$). The former accounts for the balance between precision and recall, while the latter computes area under the false positive rate and true positive rate (recall) which reflects the influence of the true negatives. We also note the multi-label accuracy $\text{Acc}$ which accounts for all binary predictions of each class. 

The most important observation on the result is that the spectrogram representation easily outperforms the audio encoder embeddings on this task, especially on the single-label prediction case (46.67\%). This offers an interesting perspective on the learned embedding content: exact note onsets and texture patterns (that are associated with the piano technique classes) seem to be overlooked by the embeddings, capturing less performance-related details compared to the lower-level spectrogram. The results demonstrate that DAC and JukeBox are the least informative audio embeddings for this task (24.44\% and 25.44\%). Audio-MAE is the best-performing audio encoder, but the single-label prediction results do not improve with fine-tuning. On the other hand, fine-tuning DAC on the solo piano data improved performance on this task by 2\%, compared with its pre-trained version. 

\begin{figure}[!h]
    \centering
    \includegraphics[width=\linewidth]{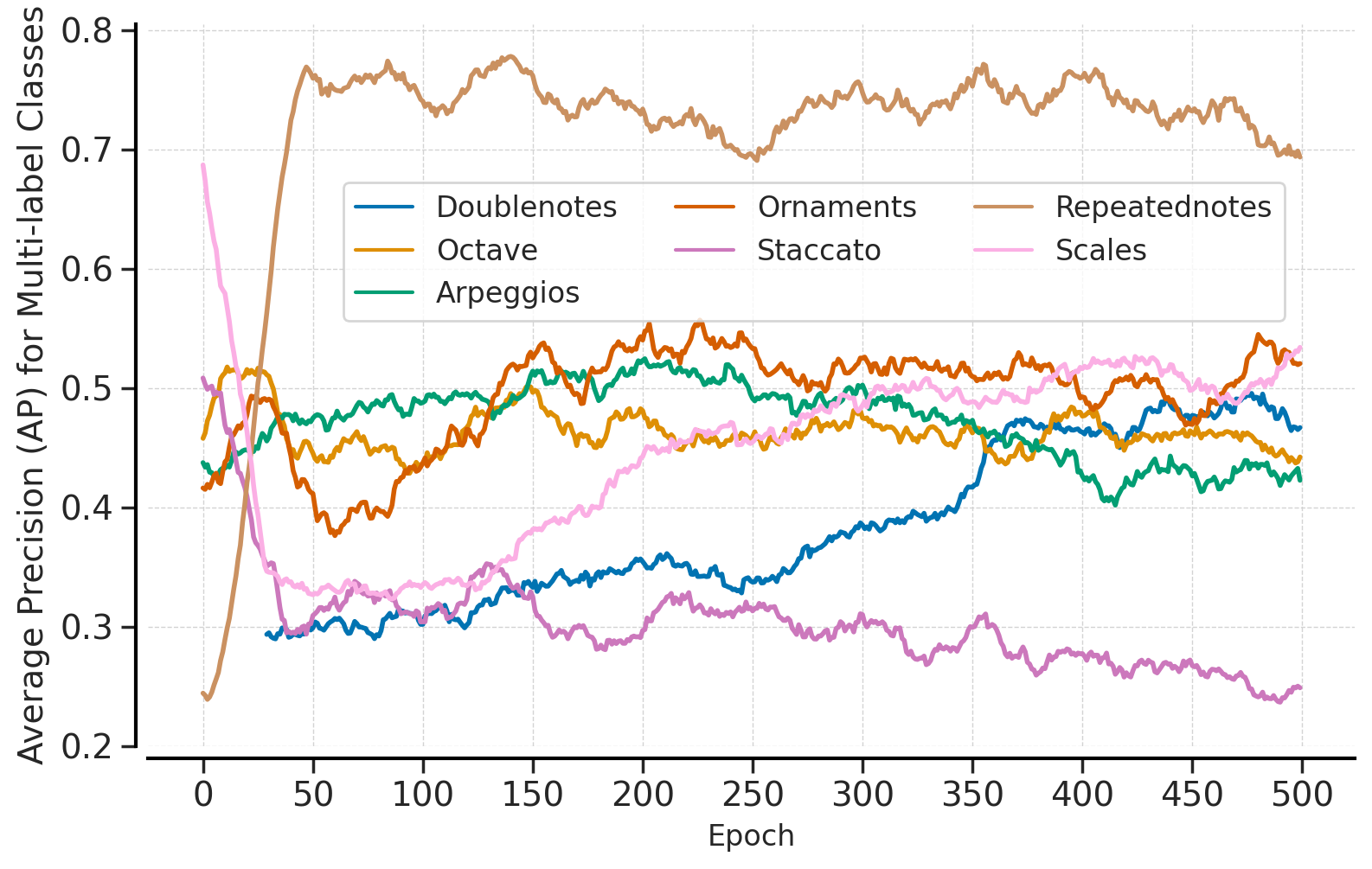}
    \caption{Average Precision for each class over epochs in multi-class prediction, from spectrogram representation.}
    \vspace{-2em}
    \label{fig:ap_techniques}
\end{figure}

To gain a better understanding of the identified techniques we observe the class-wise $mAP$ from the best-performing representation of spectrogram. As depicted in Figure~\ref{fig:ap_techniques}, Repeated Notes emerge as the most accurately identified technique. Conversely, the Staccato class exhibits a decline in performance throughout the training, hinting at a potential acoustic overlap with Repeated Notes, as suggested by prior research \cite{Bresin2000Articulation545}. Meanwhile, the precision for other techniques shows consistent improvement during training, achieving 40\% to 60\% even in more distinct technique categories like Scales and Arpeggios. However, with the highest accuracy for single-label 7-way prediction being 46.67\%, it is clear that the model's ability to pinpoint techniques could be further refined, especially considering these are discernible to the human ear.

\section{Conclusion}



Our research aimed to extend the capabilities of audio encoding models to the domain of solo piano performance understanding. Through this effort, we addressed tasks such as expertise ranking, difficulty estimation, and solo piano technique detection. The study introduced the Pianism-Labelling Dataset (PLD) and utilized a range of pre-trained audio encoders for evaluation. The curated set of performance-related attribute labeling can be contributed to multi-task learning or contrastive learning in the future. 

Our results, with the highest accuracy of 93.6\% in expertise ranking, suggest that models like Audio-MAE hold promise for assessing aspects of musical performance, while the codified representations such as DAC or Jukebox struggle with capturing performance nuances. However, the studies on difficulty and especially techniques suggest the limitations of current pre-trained representations in capturing pianistic textures and patterns, as they fail to outperform the spectrogram baseline, prompting for the design of a performance-oriented audio representation. Meanwhile, the case study on the Chopin Piano Competition via transferring the assessment objective confirmed that we are still far from capturing the nuances of top-level human performance. 


\section{Acknowledgement}

This work is supported by the UKRI Centre for Doctoral Training in Artificial Intelligence and Music, funded by UK Research and Innovation [grant number EP/S022694/1], and the Engineering and Physical Sciences Research Council [grant number EP/T518086/1]. We are grateful for pianist Yan Zhou for helping with data verification. 

\bibliography{ref}

\begin{thebibliography}{10}
\providecommand{\url}[1]{#1}
\csname url@samestyle\endcsname
\providecommand{\newblock}{\relax}
\providecommand{\bibinfo}[2]{#2}
\providecommand{\BIBentrySTDinterwordspacing}{\spaceskip=0pt\relax}
\providecommand{\BIBentryALTinterwordstretchfactor}{4}
\providecommand{\BIBentryALTinterwordspacing}{\spaceskip=\fontdimen2\font plus
\BIBentryALTinterwordstretchfactor\fontdimen3\font minus \fontdimen4\font\relax}
\providecommand{\BIBforeignlanguage}[2]{{%
\expandafter\ifx\csname l@#1\endcsname\relax
\typeout{** WARNING: IEEEtran.bst: No hyphenation pattern has been}%
\typeout{** loaded for the language `#1'. Using the pattern for}%
\typeout{** the default language instead.}%
\else
\language=\csname l@#1\endcsname
\fi
#2}}
\providecommand{\BIBdecl}{\relax}
\BIBdecl

\bibitem{Korzeniowski2017EndNetwork}
F.~Korzeniowski and G.~Widmer, ``End-to-end musical key estimation using a convolutional neural network,'' in \emph{25th European Signal Processing Conference, EUSIPCO}, 2017.

\bibitem{Pons2018End-to-endScale}
J.~Pons, O.~Nieto, M.~Prockup, E.~Schmidt, A.~Ehmann, and X.~Serra, ``End-to-end learning for music audio tagging at scale,'' \emph{Proceedings of the 19th International Society for Music Information Retrieval Conference (ISMIR)}, pp. 637--644, 2018.

\bibitem{FloresGarcia2021LeveragingRecognition}
H.~{Flores Garcia}, A.~Aguilar, E.~Manilow, and B.~Pardo, ``Leveraging hierarchical structures for few-shot musical instrument recognition,'' in \emph{Proceedings of the 22th International Society for Music Information Retrieval Conference, ISMIR 2021}, 2021.

\bibitem{Castellon2021CodifiedRetrieval}
R.~Castellon, C.~Donahue, and P.~Liang, ``Codified audio language modeling learns useful representations for music information retrieval,'' \emph{Proceedings of the 22th International Society for Music Information Retrieval Conference, ISMIR 2021}, 2021.

\bibitem{McCallum2022SupervisedUnderstanding}
M.~C. McCallum, F.~Korzeniowski, S.~Oramas, F.~Gouyon, and A.~F. Ehmann, ``Supervised and unsupervised learning of audio representations for music understanding,'' in \emph{Proceeding of the 21t International Society on Music Information Retrieval (ISMIR)}, 2022.

\bibitem{Gardner2024LLarkMusic}
J.~Gardner, S.~Durand, D.~Stoller, and R.~M. Bittner, ``{LLark}: A multimodal foundation model for music,'' in \emph{Proceedings of the International Conference on Machine Learning (ICML)}, 2024.

\bibitem{Doh2023LP-MusicCapsCaptioning}
S.~Doh, K.~Choi, J.~Lee, and J.~Nam, ``{LP-MusicCaps}: {LLM}-based pseudo music captioning,'' in \emph{Proceeding of the 24th International Society on Music Information Retrieval (ISMIR)}, 2023.

\bibitem{Palmer1997MusicPerformance}
C.~Palmer, ``Music performance,'' \emph{Annual Review of Psychology}, vol.~48, 1997.

\bibitem{Ramoneda2022ScoreFingering}
P.~Ramoneda, N.~{Can Tamer}, V.~Eremenko, X.~Serra, and M.~Miron, ``Score difficulty analysis for piano performance education based on fingering,'' in \emph{Proceeding of the IEEE International Conference on Acoustics, Speech and Signal Processing (ICASSP)}, 2022.

\bibitem{Parmar2021PianoAssessment}
P.~Parmar, J.~Reddy, and B.~Morris, ``Piano skills assessment,'' in \emph{IEEE 23th International Workshop on Multimedia Signal Processing (MMSP)}, 2021.

\bibitem{Wang2021Audio-basedMechanism}
W.~Wang, J.~Pan, H.~Yi, Z.~Song, and M.~Li, ``Audio-based piano performance evaluation for beginners with convolutional neural network and attention mechanism,'' \emph{IEEE/ACM Transactions on Audio Speech and Language Processing}, vol.~29, pp. 1119--1133, 2021.

\bibitem{Seshadri2021ImprovingLearning}
P.~Seshadri and A.~Lerch, ``Improving music performance assessment with contrastive learning,'' in \emph{Proceedings of the 22nd International Society for Music Information Retrieval Conference (ISMIR)}, 2021.

\bibitem{Huang2019AutomaticExercises}
J.~Huang and A.~Lerch, ``Automatic assessment of sight-reading exercises,'' in \emph{Proceedings of the 20th International Society for Music Information Retrieval Conference (ISMIR)}, 2019.

\bibitem{Morsi2023SoundsPerformances}
A.~Morsi, K.~Tatsumi, A.~Maezawa, T.~Fujishima, and X.~Serra, ``Sounds {Out} of {Pl{\"{a}}ce}? {S}core-independent detection of conspicuous mistakes in piano performances,'' in \emph{Proceeding of the 24th International Society on Music Information Retrieval (ISMIR)}, 2023.

\bibitem{Kim2022OverviewContext}
H.~Kim, P.~Ramoneda, M.~Miron, and X.~Serra, ``An overview of automatic piano performance assessment within the music education context,'' \emph{International Conference on Computer Supported Education, CSEDU - Proceedings}, vol.~1, 2022.

\bibitem{Morsi2024SimulatingLearning}
A.~Morsi, H.~Zhang, A.~Maezawa, S.~Dixon, and X.~Serra, ``Simulating piano performance mistakes for music learning,'' in \emph{Proceedings of the Sound and Music Computing Conference (SMC)}, 2024.

\bibitem{Lerch2020AnAnalysis}
A.~Lerch, C.~Arthur, A.~Pati, and S.~Gururani, ``An interdisciplinary review of music performance analysis,'' \emph{Transactions of the International Society for Music Information Retrieval}, vol.~3, no.~1, pp. 221--245, 2020.

\bibitem{Zhang2023SymbolicEvaluation}
H.~Zhang, E.~Karystinaios, S.~Dixon, G.~Widmer, and C.~E. Cancino-Chac{\'{o}}n, ``Symbolic music representations for classification tasks: A systematic evaluation,'' in \emph{Proceeding of the 24th International Society on Music Information Retrieval (ISMIR)}, Milan, Italy, 2023.

\bibitem{CancinoChacon2018ComputationalModels}
C.~E. Cancino-Chac{\'{o}}n, ``Computational modeling of expressive music performance with linear and non-linear basis function models,'' Ph.D. dissertation, Johannes Kepler University Linz, 2018.

\bibitem{Li2023MutimodalAssistants}
C.~Li, Z.~Gan, Z.~Yang, J.~Yang, L.~Li, L.~Wang, and J.~Gao, ``Multimodal foundation models: From specialists to general-purpose assistants,'' \emph{arXiv preprint arXiv:2309.10020}, 2023.

\bibitem{Won2024FoundationalInformatics}
M.~Won, Y.-n. Hung, and D.~Le, ``A foundation model for music informatics,'' in \emph{2024 IEEE International Conference on Acoustics, Speech and Signal Processing (ICASSP)}, 2024.

\bibitem{Li2023MERTTraining}
Y.~Li, R.~Yuan, G.~Zhang, Y.~Ma, X.~Chen, H.~Yin, C.~Xiao, C.~Lin, A.~Ragni, E.~Benetos, N.~Gyenge, R.~Dannenberg, R.~Liu, W.~Chen, G.~Xia, Y.~Shi, W.~Huang, Z.~Wang, Y.~Guo, and J.~Fu, ``{MERT}: Acoustic music understanding model with large-scale self-supervised training,'' in \emph{Proceedings of the International Conference on Learning Representations (ICLR)}, 2024.

\bibitem{Kumar2023High-FidelityRVQGAN}
R.~Kumar, P.~Seetharaman, A.~Luebs, I.~Kumar, and K.~Kumar, ``High-fidelity audio compression with improved {RVQGAN},'' in \emph{Proceedings of the Conference on Neural Information Processing Systems (Neurips)}, 2023.

\bibitem{Hu2022MaskedListen}
P.-y.~H. Hu, X.~Juncheng, C.~Feichtenhofer, and M.~Ai, ``Masked autoencoders that listen,'' in \emph{Proceedings of the Conference on Neural Information Processing Systems (NeurIPS)}, 2022.

\bibitem{Kosta2018DynamicsScore}
K.~Kosta, O.~F. Bandtlow, and E.~Chew, ``Dynamics and relativity: Practical implications of dynamic markings in the score,'' \emph{Journal of New Music Research}, vol.~47, no.~5, pp. 438--461, 2018.

\bibitem{Kim2023DiffVelModel}
H.~Kim and X.~Serra, ``{DiffVel} : Note-level midi velocity estimation for piano performance by a double conditioned diffusion model,'' in \emph{16th International Symposium on Computer Music Multidisciplinary Research (CMMR)}, Tokyo, Japan, 2023.

\bibitem{Grachten2009Phase-planeTiming}
M.~Grachten, W.~Goebl, S.~Flossmann, and G.~Widmer, ``Phase-plane representation and visualization of gestural structure in expressive timing,'' \emph{Journal of New Music Research}, vol.~38, no.~2, pp. 183--195, Jun. 2009.

\bibitem{Shi2021ComputationalMazurkas}
Z.~Shi, ``Computational analysis and modeling of expressive timing in chopin mazurkas,'' in \emph{Proceedings of the 22nd International Society for Music Information Retrieval Conference (ISMIR)}, 2021.

\bibitem{Rafee2021PerformerModels}
S.~R.~M. Rafee, G.~Fazekas, and G.~A. Wiggins, ``Performer identification from symbolic representation of music using statistical models,'' in \emph{Proceedings of the International Computer Music Conference (ICMC)}, 2021.

\bibitem{Zhao2021ViolinistFeatures}
Y.~Zhao, C.~Wang, G.~Fazekas, E.~Benetos, and M.~Sandler, ``Violinist identification based on vibrato features,'' in \emph{European Signal Processing Conference}, 2021.

\bibitem{Sales2023InteractiveEngagement}
C.~Sales, P.~Wang, and Y.~Jiang, ``An interactive tool for exploring score-aligned performances: Opportunities for enhanced music engagement,'' \emph{ACM International Conference Proceeding Series}, pp. 30--33, 2023.

\bibitem{Zhang2021LearnAssessment}
H.~Zhang, Y.~Jiang, T.~Jiang, and P.~Hu, ``Learn by referencing: Towards deep metric learning for singing assessment,'' in \emph{Proceedings of the 22nd International Society for Music Information Retrieval Conference (ISMIR)}, 2021.

\bibitem{Matsubara2021CROCUSCritiques}
M.~Matsubara, R.~Kagawa, T.~Hirano, and I.~Tsuji, ``{CROCUS}: Dataset of musical performance critiques,'' in \emph{In Proceedings of the International Symposium on Computer Music Multidisciplinary Research (CMMR)}, 2021.

\bibitem{Jiang2023ExpertFeedback}
Y.~Jiang, ``Expert and novice evaluations of piano performances : Criteria for computer-aided feedback,'' in \emph{Proceeding of the 24th International Society on Music Information Retrieval (ISMIR)}, 2023.

\bibitem{Ramoneda2023CombiningClassification}
P.~Ramoneda, D.~Jeong, V.~Eremenko, N.~C. Tamer, M.~Miron, and X.~Serra, ``Combining piano performance dimensions for score difficulty classification,'' \emph{Expert Systems with Applications}, 2024.

\bibitem{Wang2020PlayingScattering}
C.~Wang, V.~Lostanlen, E.~Benetos, and E.~Chew, ``Playing technique recognition by joint time frequency scattering,'' in \emph{IEEE International Conference on Acoustics, Speech and Signal Processing (ICASSP)}, 2020.

\bibitem{Srivatsan2022ChecklistPrediction}
N.~Srivatsan and T.~Berg-kirkpatrick, ``Checklist models for improved output fluency in piano fingering prediction,'' in \emph{Proceeding of the 23rd International Society on Music Information Retrieval (ISMIR)}, 2022.

\bibitem{Hall2011CalibratingStudents}
C.~V. Hall and J.~T. O'Donnell, ``Calibrating a bowing checker for violin students,'' \emph{Journal of Music, Technology \& Education}, vol.~3, no. 2-3, pp. 125--139, 2011.

\bibitem{Ramoneda2024CanDataset}
P.~Ramoneda, M.~Lee, D.~Jeong, J.~J. Valero-Mas, and X.~Serra, ``Can audio reveal music performance difficulty? insights from the piano syllabus dataset,'' pp. 1--13, 2024.

\bibitem{Borsos2023AudioLMGeneration}
Z.~Borsos, R.~Marinier, D.~Vincent, E.~Kharitonov, O.~Pietquin, M.~Sharifi, D.~Roblek, O.~Teboul, D.~Grangier, M.~Tagliasacchi, and N.~Zeghidour, ``Audiolm: A language modeling approach to audio generation,'' \emph{IEEE/ACM Transactions on Audio Speech and Language Processing}, 2023.

\bibitem{Agostinelli2023MusicLMText}
A.~Agostinelli, T.~I. Denk, Z.~Borsos, J.~Engel, M.~Verzetti, A.~Caillon, Q.~Huang, A.~Jansen, A.~Roberts, M.~Tagliasacchi, M.~Sharifi, N.~Zeghidour, and C.~Frank, ``Musiclm: Generating music from text,'' \emph{arXiv preprint arXiv:2301.11325}, 2023.

\bibitem{Dhariwal2020JukeboxMusic}
P.~Dhariwal, H.~Jun, C.~Payne, J.~W. Kim, A.~Radford, and I.~Sutskever, ``Jukebox: A generative model for music,'' \emph{Computing Research Repository (CoRR)}, 2020.

\bibitem{Manilow2022SourceModels}
E.~Manilow, P.~O'Reilly, P.~Seetharaman, and B.~Pardo, ``Source separation by steering pretrained music models,'' in \emph{ICASSP, IEEE International Conference on Acoustics, Speech and Signal Processing - Proceedings}, vol. 2022-May, 2022, pp. 126--130.

\bibitem{Liu2024MusicCaptioning}
S.~Liu, A.~S. Hussain, C.~Sun, and Y.~Shan, ``Music understanding {LLaMA}: Advancing text-to-music generation with question answering and captioning,'' in \emph{Proceeding of the IEEE International Conference on Acoustics, Speech and Signal Processing (ICASSP)}, 2024.

\bibitem{Liang2023AcousticCapabilities}
J.~Liang, ``Acoustic prompt tuning: Empowering large language models with audition capabilities,'' \emph{arXiv preprint arXiv:2312.00249}, 2023.

\bibitem{Garcia2023VampNetModeling}
H.~F. Garcia, P.~Seetharaman, R.~Kumar, and B.~Pardo, ``{VampNet}: Music generation via masked acoustic token modeling,'' in \emph{Proceeding of the 24th International Society on Music Information Retrieval (ISMIR)}, 2023.

\bibitem{Puvvada2024DiscreteRecognition}
K.~C. Puvvada, N.~R. Koluguri, K.~Dhawan, J.~Balam, and B.~Ginsburg, ``Discrete audio representation as an alternative to mel-spectrograms for speaker and speech recognition,'' in \emph{Proceeding of the IEEE International Conference on Acoustics, Speech and Signal Processing (ICASSP)}, 2024.

\bibitem{Hawthorne2018EnablingDataset}
C.~Hawthorne, A.~Stasyuk, A.~Roberts, I.~Simon, C.~Z.~A. Huang, S.~Dieleman, E.~Elsen, J.~Engel, and D.~Eck, ``Enabling factorized piano music modeling and generation with the {Maestro} dataset,'' in \emph{Proceedings of the International Conference on Learning Representations (ICLR)}, 2019, pp. 1--12.

\bibitem{Zhang2022ATEPPPerformance}
H.~Zhang, J.~Tang, S.~Rafee, S.~Dixon, and G.~Fazekas, ``{ATEPP}: A dataset of automatically transcribed expressive piano performance,'' in \emph{Proceedings of the International Society for Music Information Retrieval Conference (ISMIR)}, Bengaluru, India, 2022.

\bibitem{Konz2011SaarlandData}
V.~Konz, W.~Bogler, and V.~Arifi-M, ``Saarland music data,'' \emph{Late-Breaking and Demo Session of the International Society on Music Information Retrieval (ISMIR)}, 2011.

\bibitem{Sandor1981OnExpression}
G.~Sandor, \emph{On Piano Playing: Motion, Emotion and Expression}, 1981.

\bibitem{Tsay2013SightPerformance}
C.-J. Tsay, ``Sight over sound in the judgment of music performance,'' \emph{Proceedings of the National Academy of Sciences}, 2013.

\bibitem{Bresin2000Articulation545}
R.~Bresin and G.~{Umberto Battel}, ``Articulation strategies in expressive piano performance analysis of legato, staccato, and repeated notes in performances of the {Andante} movement of {Mozart}'s sonata in g major (k.545),'' \emph{Journal of New Music Research}, vol.~29, no.~3, pp. 211--224, 2000.

\end{thebibliography}

%
%
%
%
%

\end{document}